\documentclass[cameraready]{Interspeech}

\usepackage{arydshln, subcaption}

\title{Fast Speech Foundation Model Distillation Using Interleaved Stacking}

\author[affiliation={1}, orcid=0000-0003-1723-6776]{Eungbeom}{Kim}
\author[affiliation={1,2,3}, orcid=0000-0002-4210-0312]{Kyogu}{Lee}

\address{
    $^1$IPAI, $^2$AIIS, $^3$Dept. of Intelligence and Information\\
    Seoul National University, Seoul, Republic of Korea
}

\email{eb.kim@snu.ac.kr, kglee@snu.ac.kr}


\usepackage{comment}

\begin{document}

\maketitle

\begin{abstract}
    Distilling a large speech foundation model (SFM) into an efficient student model has been successfully applied to low-resource environments.
    Although distillation reduces inference latency, it requires an additional student model training.
    However, the training efficiency of SFM distillation remains underexplored.
    In this work, we explore training acceleration of SFM distillation to speed up model deployment.
    We examine the potential of stacking, in which the model depth is progressively increased through training until the target model depth is reached.
    While existing stacking methods improve training speed, they suffer from performance degradation.
    To handle this limitation, we propose interleaved stacking, a novel stacking method that consistently preserves layer position throughout the stacking process.
    This property is particularly critical in SFMs, in which each layer encodes distinct layer-specific knowledge.
    We validate the effectiveness of the proposed method on SUPERB.
\end{abstract}

\section{Introduction}
Knowledge distillation (KD)~\cite{Hinton06} is one of the most popular approaches for deploying deep learning models in real-world environments.
By training an efficient student model with low computational requirements, KD avoids the expensive cost of large-scale models.
In speech processing, KD has demonstrated strong impact for speech foundation models (SFMs).
For instance, DistilHuBERT~\cite{distilhubert} reduces the depth of the Transformer model~\cite{transformer} to two layers while preserving the width of the teacher SFM.
Likewise, shallow-and-wide student architectures have been widely adopted for their remarkable efficiency, but suffer from performance degradation on complex speech processing tasks such as automatic speech recognition (ASR), phoneme recognition (PR), and slot filling (SF).
As an alternative, \cite{deepwide, fithubert, armhubert} mitigate performance degradation by employing deep-and-narrow architectures for the student model.
While deep-and-narrow architectures achieve competitive performance under memory constraints, \cite{lin2022smaller} observes that they induce limited speed improvement compared to shallow-and-wide architectures.
Although this is expected given that shallow-and-wide architectures are more suited to parallel computing, this leads to delayed training of deep-and-narrow student models.

\begin{figure}[t]
    \centering
    \includegraphics[width=\linewidth]{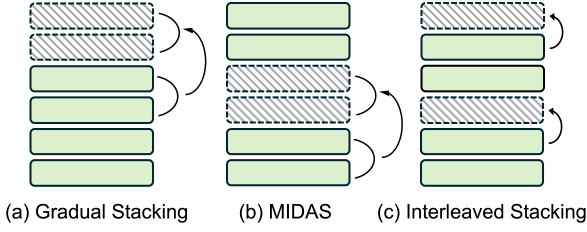}
    \caption{Illustration of various stacking approaches: (a) gradual stacking, (b) MIDAS, and (c) interleaved stacking (ours).}
    \label{fig:stacking}
\end{figure}

To address this issue, we focus on training acceleration via stacking-based stagewise training, in which the model is progressively extended through training stages.
At the end of each stage, the model depth is increased for the next stage, and this process is repeated until the target model depth is reached.
By doing so, the total training cost is reduced because early stages only train low-cost shallow architectures.
However, since there is a mismatch between the target model and the early stage models, stacking can lead to performance degradation of the target model.
In this context, various stacking strategies have been explored to prevent performance degradation, as shown in Figure~\ref{fig:stacking}.
For instance, gradual stacking~\cite{gradstack} copies the last $K$ layers and stacks them on top of the model at the end of each stage, successfully boosting training speed while maintaining competitive downstream performance on language modeling tasks.
Instead of selecting the last $K$ layers, MIDAS~\cite{midas} utilizes the middle layers and shows that stacking can even surpass conventional models on language-based reasoning tasks under a well-defined stacking strategy.

Despite the potential of stacking for training acceleration, stacking for SFM distillation remains uninvestigated.
Therefore, our primary goal in this work is to examine the effectiveness of stacking approaches for SFM distillation and to design a stacking approach that reduces training costs without sacrificing the performance of the model.
In particular, we focus on the inconsistent layer positions across training stages in existing stacking approaches.
For instance, in gradual stacking, the late layers are relocated to earlier positions across training stages, because the copied layers are repeatedly stacked on top of the model.
Since the SFMs are known to exhibit layer-specific knowledge~\cite{pasad2021layer, wavlm, el2025comprehensive}, we argue that this inconsistency is undesirable for SFM distillation.
Instead, we introduce interleaved stacking, in which each copied layer is inserted right after its original layer, as shown in Figure~\ref{fig:stacking}.
By doing so, the early and late layers in the initial stage remain at the proportionally early and late positions, respectively, in subsequent stages.
This is further inspired by the prior studies~\cite{nguyen2021do, jiang2025tracing} confirming that adjacent Transformer layers behave similarly, which also holds for SFMs~\cite{el2024speech}.
Furthermore, due to this advantage, intermediate-level KD losses are naturally integrated into the proposed method, which is critical for SFM distillation~\cite{distilhubert, fithubert}.

We evaluate our training acceleration method for SFM distillation on various speech processing downstream tasks using the SUPERB benchmark~\cite{superb}.
First, we verify the effectiveness of existing stacking approaches for training acceleration in terms of reduced training cost requirements, although there is performance degradation on downstream tasks.
Building upon this, we demonstrate that the proposed method successfully mitigates performance degradation of the existing stacking baselines across content, semantics, and speaker-based speech processing tasks under multiple scheduling strategies.
Furthermore, we verify the seamless compatibility of the proposed stacking method with the intermediate-level KD losses, achieving extra performance gains.

\begin{figure}[t]
    \centering
    \includegraphics[width=\linewidth]{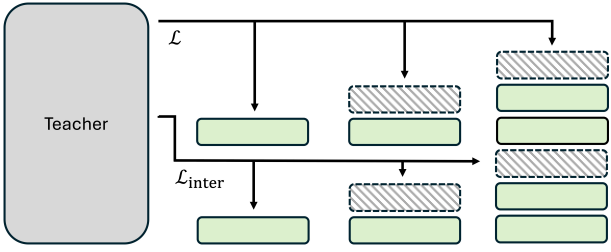}
    \caption{Illustration of the proposed training framework with interleaved stacking. The total loss function consists of the output-level KD loss and the intermediate-level KD loss.}
    \label{fig:loss}
\end{figure}

\section{Background}
Knowledge distillation trains a small $N$-layer student model $f=f_N\circ\dots\circ f_1=f_{N:1}$ using the knowledge of a large $M$-layer teacher model for model compression.
Here, our goal is to accelerate student model training via stagewise training using a stacking approach.
We consider $B$-stage training for stacking.
At the initial stage, a shallow $K$-layer model is initialized where $K=N/B$, and at the end of the training stage, $K$ layers are copied and appended to the previous model for the next stage.
For instance, after the $K$-layer model is initialized and trained in the initial stage, a $2K$-layer model is initialized based on the previous stage model and then trained in the second stage.
This is repeated until the $B$-th stage, where the model reaches to the target depth $N$.
That is, the $b$-th stage model has $bK$ layers, and the final $B$-th stage model has the full $BK=N$ layers.
Since shallow models are trained in the early stages, the overall training cost is reduced by the stagewise training approach.

For stagewise training, several stacking approaches have been investigated in the prior studies.
\textbf{Gradual stacking}~\cite{gradstack} is a popular stacking baseline in which the last $K$ layers are copied and stacked on top of the model.
That is, the $(b+1)$-th stage model is initialized by $f_{bK:(b-1)K+1}\circ f_{bK:(b-1)K+1}\circ f_{(b-1)K:1}$, where $f_{bK:1}$ is the $b$-th stage model.
\textbf{MIDAS}~\cite{midas} selects an intermediate block of $K$ layers for stacking; thus, the $(b+1)$-th stage model is initialized by $f_{bK:iK+1}\circ f_{iK:(i-1)K+1}\circ f_{iK:(i-1)K+1}\circ f_{(i-1)K:1}$, where $i=\lceil b/2\rceil$ denotes the intermediate block.
In this work, we explore the effectiveness of stacking for fast SFM distillation for the first time, to the best of our knowledge.

\section{Interleaved Stacking}
While stacking reduces training cost, we find that it negatively impacts downstream performance.
Thus, we propose a novel stacking method, \textbf{interleaved stacking}, for SFM distillation.
Instead of copying and stacking the consecutive $K$ layers in the model as in the existing approaches, in the $b$-th stage we select every $b$-th layer to copy $K$ layers in total and insert each copied layer after its original layer.
That is, the $(b+1)$-th stage model is initialized by copying and stacking the layer $f_{bk}$ from the $b$-th stage model, as $f_{bk}=f_{bk}\circ f_{bk}$ for $k=1,\dots,K$, which is also depicted in Figure~\ref{fig:stacking} (c).

By doing so, we preserve layer position consistency, in contrast to existing methods such as gradual stacking, where the last layer of the initial stage model is repositioned to the middle of the model at the next stage.
This is critical for SFM distillation in that such positional inconsistency hinders the distillation of layer-specific knowledge in SFMs~\cite{pasad2021layer, wavlm, el2025comprehensive}.
Moreover, this layer position consistency of the proposed method makes the distillation naturally compatible with the intermediate-level KD losses, which are a key component of SFM distillation~\cite{distilhubert, fithubert, deepwide}.
Additionally, by locating each copied layer at the subsequent position of its original layer, the copied layers are placed in similar position to their original layers, which is motivated by the prior studies~\cite{nguyen2021do, el2024speech, jiang2025tracing} that empirically confirm the similarity between adjacent layers.

Next, we combine the proposed stacking approach with SFM distillation.
We consider a frame-level MSE loss function between the teacher SFM output $y^T\in\mathbb{R}^{F\times D^T}$ and the student SFM output $y\in\mathbb{R}^{F\times D}$ to train the student model following~\cite{menon2021statistical, fithubert, armhubert}, where $F$ denotes the total number of frames, and $D^T$ and $D$ denotes the teacher and student output dimensions, respectively.
To project the $D$-dimensional student representation to the teacher dimension $D^T$, we utilize a simple linear projection layer $\text{proj}(\cdot)$.
The output-level KD loss function $\mathcal{L}$ is formulated as follows: $\mathcal{L}=\text{MSE}(y^T,\text{proj}(y))$.
We also adopt the intermediate-level KD loss using the layer-to-layer MSE loss inspired by~\cite{Romero15-iclr, fithubert, deepwide}.
However, the student model has reduced depth in stacking approaches except the last stage.
Thus, we set the $(K-1)$ intermediate-level KD losses for the $K$-layer initial stage student model and maintain these throughout training.
For the target representations, every $M/K$-th teacher intermediate representation is utilized except the final output as follows: $\mathcal{L}_{\text{inter}}=\sum^{K-1}_{k=1}\text{MSE}(y^T_{kM/K}, \text{proj}_k(y_{bk}))$, where $y^T_i$ and $y_i$ denote the $i$-th layer representations of the teacher model and the student model, respectively, and $\text{proj}_i(\cdot)$ is the $i$-th projection layer.
Note that the target layer indices of the teacher model for the intermediate-level KD loss are fixed across all stages to ensure stable and fast training.
For instance, given a 12-layer teacher and student models, $B=4$ training stages, and a $3$-layer initial stage model ($K=N/B=3$), the intermediate-level KD loss is set to $\mathcal{L}_{\text{inter}}=\text{MSE}(y^T_4,y_1)+\text{MSE}(y^T_8,y_{2})$ for $b=1$ and $\mathcal{L}_{\text{inter}}=\text{MSE}(y^T_4,y_2)+\text{MSE}(y^T_8, y_4)$ for $b=2$.
Finally, the total loss function is defined as $\mathcal{L}_{\text{total}}=\mathcal{L}+w\mathcal{L}_{\text{inter}}$, where the hyperparameter $w$ controls the weight of the intermediate-level KD loss function.
The overall training process is simply illustrated in Figure~\ref{fig:loss}.

\begin{table*}[t]
  \caption{Results of the SFM distillation approaches on SUPERB PR, ASR, SF, and SID tasks. HuBERT is adopted for the teacher model in common, and the two backbone models (Full and Full L2L) trained without stacking is evaluated for experiments. For scheduling of stagewise training, equal scheduling and prop-1 scheduling are utilized for experiments. The best result of each scheduling strategy is in boldface, and the proposed method demonstrates significant performance improvement across all tasks.}
  \label{tab:main}
  \centering
  \begin{tabular}{lcccccccc}
    \toprule
    &&&& PR & ASR & \multicolumn{2}{c}{SF} & SID \\\cmidrule(lr){5-5}\cmidrule(lr){6-6}\cmidrule(lr){7-8}\cmidrule(lr){9-9}
    Model & Schedule & Speed & \#Params & PER $\downarrow$ & WER $\downarrow$ & F1 $\uparrow$ & CER $\downarrow$ & Acc $\uparrow$\\
    \midrule
    HuBERT & - & - & 94.68M & 5.41 & 6.42 & 88.53 & 25.20 & 81.42\\\cmidrule{1-9}
    DistilHuBERT~\cite{distilhubert} & - & - & 23.49M & 16.27 & 13.37 & 82.57 & 35.59 & 73.54\\
    12-L HALF~\cite{deepwide} & - & - & 26.87M & 13.09 & 11.87 & 84.49 & 32.54 & 69.11\\
    12-L HALF L2L~\cite{deepwide} & - & - & 26.87M & 10.67 & 10.96 & 86.11 & 30.93 & 69.52\\
    ARMHuBERT~\cite{armhubert} & - & - & 26.45M & 7.72 & 9.96 & 87.59 & 26.06 & 65.03\\
    DPHuBERT~\cite{dphubert} & - & - & 23.59M & 9.67 & 10.47 & 86.86 & 28.26 & 76.83\\\cmidrule{1-9}
    Full & - & $\times1.05$ & 26.87M & 10.22 & 10.33 & 83.42 & 31.62 & 73.83\\
    Full L2L & - & $\times1.00$ & 26.87M & 9.12 & 9.92 & 86.47 & 28.58 & 72.99\\\cdashline{1-9}
    GradStack & EQL & $\times1.25$ & 26.87M & 11.50 & 11.04 & 84.34 & 31.15 & 70.89\\
    MIDAS & EQL & $\times1.25$ & 26.87M & 10.75 & 10.74 & 83.43 & 32.22 & 69.94\\
    InterleaveStack (ours) & EQL & $\times1.24$ & 26.87M & \textbf{9.08} & \textbf{10.22} & \textbf{86.36} & \textbf{27.50} & \textbf{72.26}\\\cdashline{1-9}
    GradStack & PROP-1 & $\times1.17$ & 26.87M & 10.39 & 10.65 & 84.35 & 31.50 & 70.65\\
    MIDAS & PROP-1 & $\times1.17$ & 26.87M & 10.35 & 10.72 & 84.37 & 30.56 & 72.04\\
    InterleaveStack (ours) & PROP-1 & $\times1.16$ & 26.87M & \textbf{8.88} & \textbf{9.99} & \textbf{85.70} & \textbf{28.45} & \textbf{73.60}\\    
    \bottomrule
  \end{tabular}
\end{table*}

\section{Experiments}
\subsection{Experimental Setup}
For all experiments, the HuBERT base model~\cite{hubert} with 94.68M parameters is considered for the teacher model.
For the student model, we consider the 12-layer Transformer model with an output dimension of 384, a feed-forward dimension of 1536, and 8 attention heads, resulting in 26.87M parameters.
The student model is optimized with AdamW~\cite{loshchilov2018decoupled} using a learning rate of $5\times10^{-4}$, a batch size of 64, and a weight decay of $1\times10^{-4}$ for 75 epochs on the 960-hour LibriSpeech dataset~\cite{panayotov2015librispeech}.
A linear learning rate decay schedule is applied with warmup over the initial 7\% of training steps.
The trained models are evaluated on phoneme recognition (PR), automatic speech recognition (ASR), slot filling (SF), and speaker identification (SID) using the speech processing benchmark, SUPERB~\cite{superb}.
The models are frozen during downstream task training, and we follow the standard SUPERB configuration while adopting a learning rate of $5\times10^{-5}$ for SID, following prior studies~\cite{dphubert, jang2024star, dicehubert}.

For stacking approaches, stagewise training is conduced with $B=4$ stages.
Since the student model has 12 layers, $K=N/B=3$ layers are extended at each stage.
We experiment with two scheduling strategies including the equal strategy (eql), where training steps are evenly divided across all stages, and the prop-1 strategy (prop), where training steps are divided proportionally by stage index.
we measure speedup based on wall-clock time following the prior studies~\cite{gradstack, midas}.

\begin{figure}[t]
\centering
\begin{subfigure}{.33\columnwidth}
    \centering
    \includegraphics[width=\linewidth]{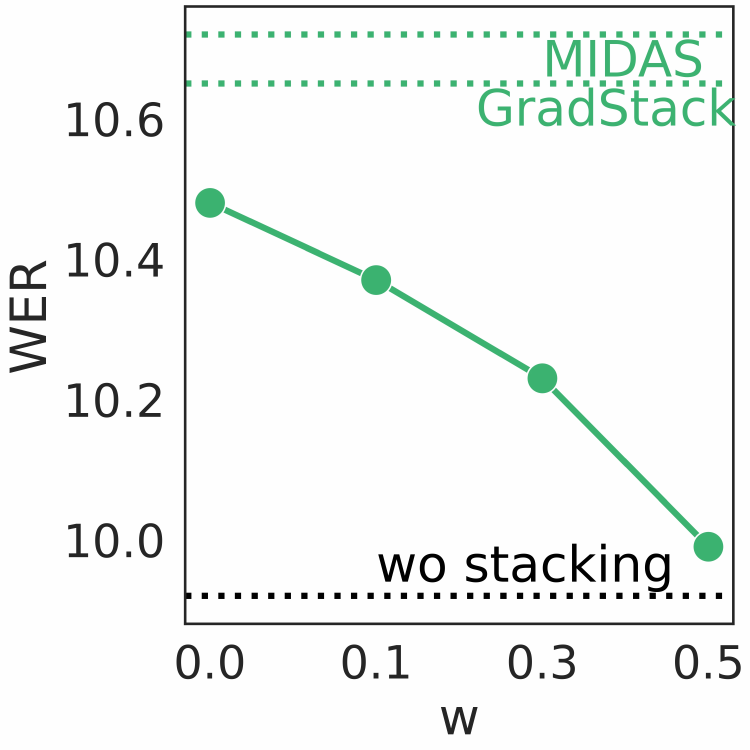}
    \caption{ASR}
    \label{fig:asr}
\end{subfigure}%
\begin{subfigure}{.33\columnwidth}
    \centering
    \includegraphics[width=\linewidth]{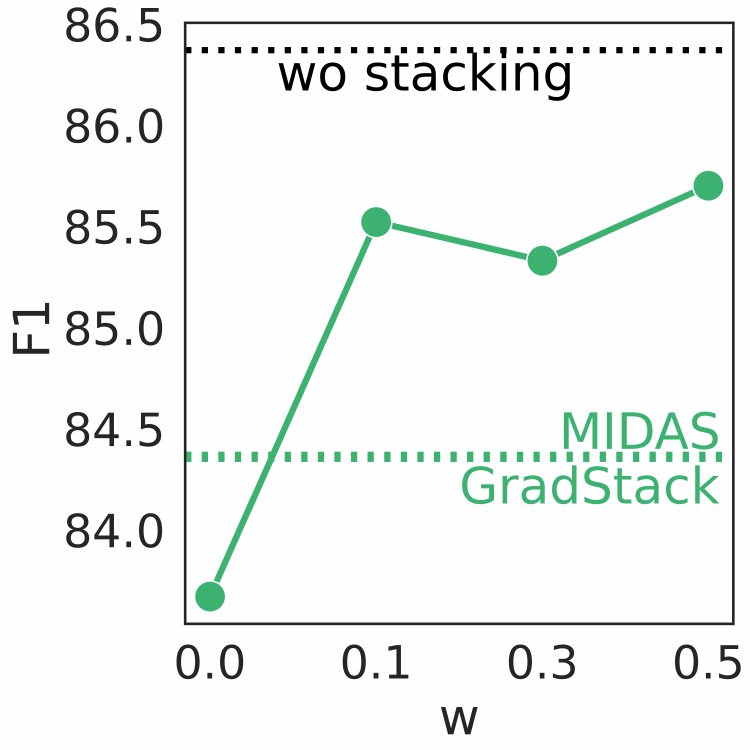}
    \caption{SF}
    \label{fig:sf}
\end{subfigure}%
\begin{subfigure}{.33\columnwidth}
    \centering
    \includegraphics[width=\linewidth]{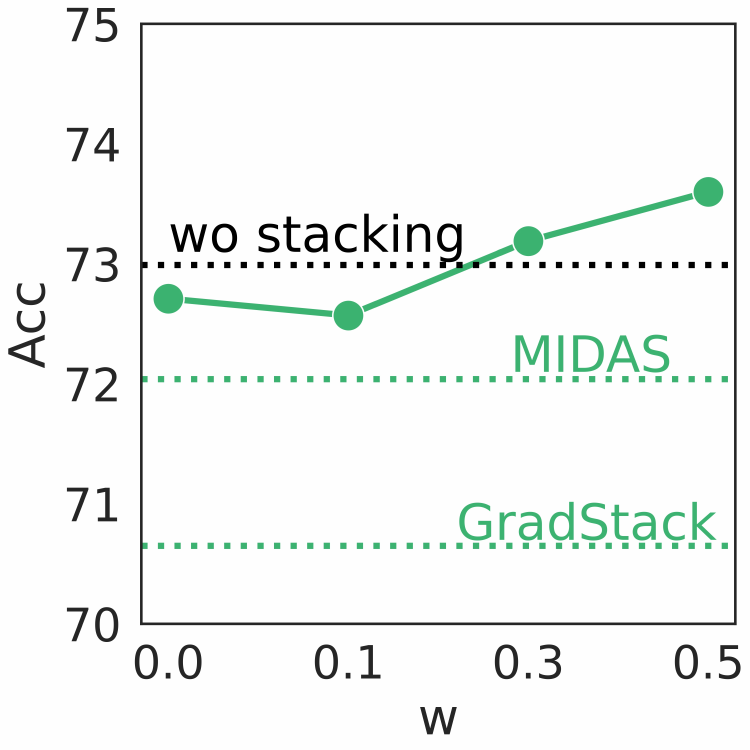}
    \caption{SID}
    \label{fig:sid}
\end{subfigure}%
\caption{
    Effect of the intermediate-level KD loss weight $w\in\{0.0,\ 0.1,\ 0.3,\ 0.5\}$ on interleaved stacking for SFM distillation. The evaluation is conducted on three tasks, ASR, SF, and SID, using the SUPERB benchmark.
}
\label{fig:w}
\end{figure}

\begin{table}[t]
  \caption{Comparison with and without the prediction style intermediate-level KD loss on gradual stacking.}
  \label{tab:intergrad}
  \centering
  \begin{tabular}{lcccc}
    \toprule
    & ASR & \multicolumn{2}{c}{SF} & SID \\\cmidrule(lr){2-2}\cmidrule(lr){3-4}\cmidrule(lr){5-5}
    Model &  WER $\downarrow$ & F1 $\uparrow$ & CER $\downarrow$ & Acc $\uparrow$\\
    \midrule
    GradStack & \textbf{10.65} & \textbf{84.35} & \textbf{31.50} & 70.65\\
    + pred-sty & 10.85 & 82.72 & 33.05 & \textbf{71.34}\\
    \bottomrule
    \end{tabular}
\end{table}

\subsection{Main Results}
Table~\ref{tab:main} presents the main results of our experiments.
To investigate the effect of stacking, we train the backbone model without stacking using the output-level KD loss (Full) and the output-level KD loss with the intermediate-level KD loss (Full L2L), following the layer-to-layer (L2L) distillation loss in \cite{deepwide}.
Overall, the proposed method demonstrates its efficiency across various tasks, successfully reducing required training time.
Our method outperforms the existing stacking baselines under both scheduling strategies by a large margin on all tasks, PR, ASR, SF, and SID.
For instance, interleaved stacking with the equal scheduling strategy achieves 9.08 PER on PR, surpassing gradual stacking achieving 11.50 PER and MIDAS achieving 10.75 PER.
As expected, the prop-1 scheduling strategy with $\times1.16$ speedup yields superior performance compared to the equal scheduling strategy with $\times1.24$ speedup except on SF.

Since our goal is to accelerate distillation while mitigating performance degradation relative to the full training framework, we also compare our method against the model trained without stacking.
Surprisingly, although the proposed method with prop-1 scheduling is trained under the reduced training cost constraint, it surpasses the full training model in terms of PER on PR, CER on SF, and accuracy on SID, with only marginal degradation such as 0.07\% increase in WER on ASR, verifying the effectiveness of the proposed method.
Furthermore, under the equal scheduling strategy, interleaved stacking outperforms the Full model trained without stacking on three out of four tasks and is competitive to the Full L2L model despite a reduced training cost.

\subsection{Intermediate-Level KD Loss}
\subsubsection{On Interleaved Stacking}
We investigate the importance of the intermediate-level KD loss $\mathcal{L}_{\text{inter}}$ by adjusting the loss weight $w\in\{0.1,\ 0.3,\ 0.5\}$.
As shown in~Figure~\ref{fig:w}, we observe that the proposed method consistently outperforms the existing methods across all $w\in\{0.1,\ 0.3,\ 0.5\}$ on SF, ASR, and SID with the prop-1 scheduling strategy.
We set $w=0.5$ for our method as a default setup because it achieves the best average results on SUPERB dev sets.

Although one of the strengths of the proposed method lies in its compatibility with the intermediate-level KD loss due to consistent layer positions, we also evaluate the case of $w=0$, which denotes that the model is trained solely on the output-level loss $\mathcal{L}$ without the intermediate-level KD loss $\mathcal{L}_{\text{inter}}$.
We find that incorporating the intermediate-level KD loss boosts performance improvement by a substantial margin, which is consistent with prior studies confirming the importance of the intermediate-level supervision in KD.
Nevertheless, it is observed that the proposed method outperforms the existing stacking baselines on ASR and SID even without the intermediate-level KD loss, demonstrating the effectiveness of the proposed method.
To further investigate this, we also analyze the effect of the intermediate-level KD loss on gradual stacking.

\subsubsection{On Gradual Stacking}
Unlike the proposed method, interleaved stacking, gradual stacking struggles to incorporate the intermediate-level KD loss.
In gradual stacking, the last $K$ layers are copied and stacked on top of the current stage model, resulting in inconsistent layer positions across training stages.
For instance, late layers from the early stages are easily repositioned to the middle or early positions as training proceeds.
For this reason, applying the layer-to-layer matching loss between the teacher and student intermediate outputs causes training instability, and we observe the loss diverges easily.
As an alternative, we adopt a prediction-style intermediate-level KD loss similar to DistilHuBER~\cite{distilhubert}, in which the final student output is used to predict multiple intermediate layer outputs of the teacher model.
Table~\ref{tab:intergrad} shows the results of the prediction-style intermediate-level KD loss applied to gradual stacking with the prop-1 scheduling strategy.
Although the prediction style intermediate-level KD loss improves accuracy on SID, it is not effective for gradual stacking overall, deteriorating performance on ASR and SF.
That is, gradual stacking is incompatible with the intermediate-level KD loss, whereas interleaved stacking successfully leverage it due to consistent layer positions across stages.

\begin{figure}[t]
\centering
\begin{subfigure}{.45\columnwidth}
    \centering
    \includegraphics[width=\linewidth]{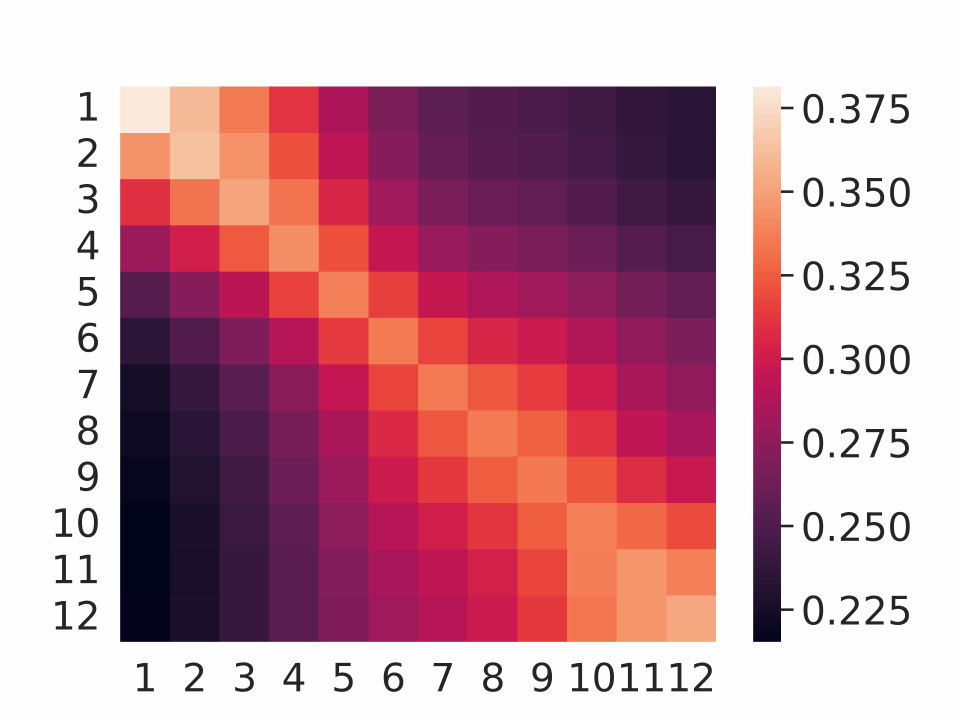}
    \caption{Full L2L}
    \label{fig:sim_full}
\end{subfigure}%
\begin{subfigure}{.45\columnwidth}
    \centering
    \includegraphics[width=\linewidth]{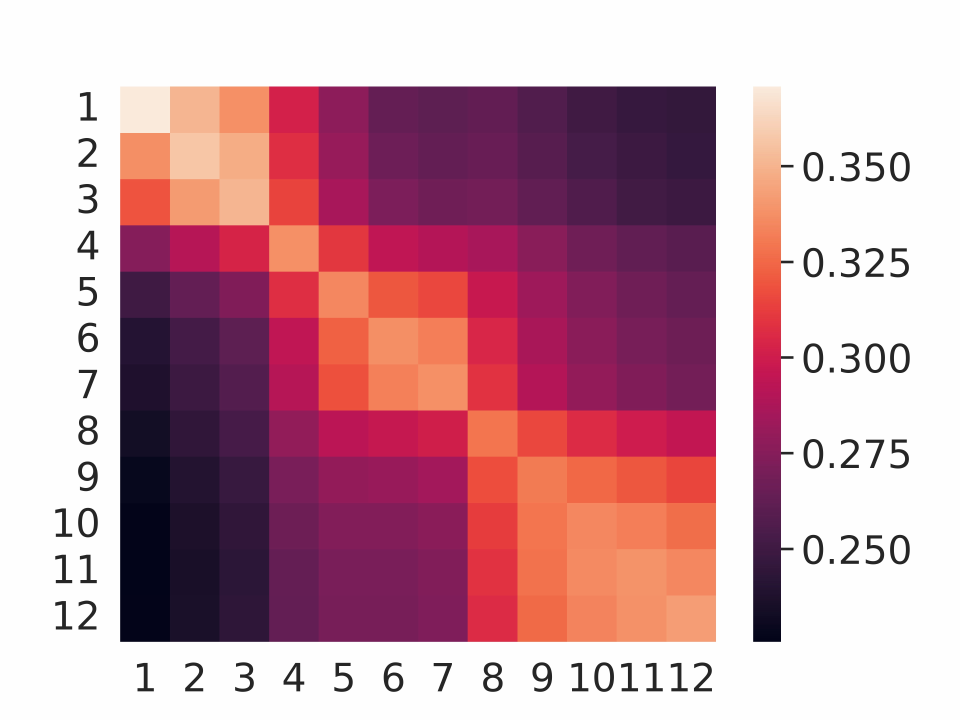}
    \caption{Interleaved Stacking}
    \label{fig:sim_interleavestack}
\end{subfigure}%
\caption{
    Cross-layer similarity of the distilled SFMs trained with (a) the conventional full training framework without stacking (Full L2L) and (b) interleaved stacking.
}
\label{fig:sim}
\end{figure}

\subsection{Layer Similarity Analysis}
We conduct an analysis of cross-layer similarity in the distilled SFMs.
Normalized cosine similarity is used to measure cross-layer similarity on the LibriSpeech dev-other dataset, and Figure~\ref{fig:sim} shows the results for the distilled models trained (a) without stacking and (b) with interleaved stacking using equal scheduling.
In the full training setting (Full L2L), we observe that the adjacent layers exhibit higher correlation than distant layers, which supports our design choice locating the copied layers at adjacent positions.
Furthermore, the early layers before the 5-th layer and the late layers after the 5-th layer seems to respectively exhibit high similarity. This aligns with the effectiveness of selecting layers at a regular interval for stacking, by which copying the similar layers are avoided, and layers with diverse knowledge are uniformly selected.

The proposed model exhibits finer block-like layer similarity structures compared to the conventionally trained model, suggesting that the role of each layer is more clearly separated.
Specifically, we observe that copied layers and their original layers, i.e., the 1st, 2nd, and 3rd layers or the 5th, 6th, and 7th layers retain high similarity after training, indicating that the copied layers serve similar roles, although the 4th layer and the 8th layer show slightly different patterns due to the position of the intermediate-level KD loss.

\section{Related Work}
\cite{deepwide} analyzes a student architecture for SFM distillation by comparing a shallow-and-wide architecture similar to~\cite{distilhubert} and a deep-and-narrow architecture.
In addition, \cite{armhubert} improves architectural efficiency of the student model utilizing parameter reusing.
Furthermore, various loss functions for SFM distillation have been investigated, including the combined L1 and cosine similarity loss, MSE loss, and self-supervised learning-based loss~\cite{distilhubert, fithubert, dicehubert}.
Similarly, \cite{distilhubert, fithubert, deepwide} demonstrate the importance of appropriate intermediate-level KD losses.
Likewise, knowledge distillation~\cite{Hinton06} has widely been studied for SFM compression supported by its effectiveness.
However, the training efficiency of SFM distillation has not been systematically studied, although training cost is also an essential factor in real-world deployment.
To address this issue, this work focuses on stagewise training via stacking, exploring various stacking approaches for SFM distillation.

\section{Conclusion}
In this work, we study a training acceleration approach based on stacking for speech foundation model (SFM) distillation, which has been underexplored despite its practical importance.
We advance the understanding of stacking-based training acceleration for SFM distillation and introduce a novel stacking method, interleaved stacking.
Unlike existing approaches, the copied layers are inserted adjacent to their original layers while stacking, resulting in consistent layer positions across multiple training stages.
Consequently, layer-specific knowledge is preserved throughout training, and the intermediate-level KD loss is easily incorporated with the proposed method.
We experimentally verify the effectiveness of the proposed method under multiple scheduling strategies on SUPERB, demonstrating that the proposed method outperforms existing baselines by a large margin.

\section{Acknowledgments}
This work was partly supported by the National Research Foundation of Korea(NRF) grant funded by the Korea government(MSIT) [No.RS-2025-24683892, 50\%] and Institute of Information \& communications Technology Planning \& Evaluation (IITP) grant funded by the Korea government(MSIT) [No.RS-2022-II220320, Artificial intelligence research about cross-modal dialogue modeling for one-on-one multi-modal interactions, 45\%] and [No.RS-2021-II211343, Artificial Intelligence Graduate School Program (Seoul National University), 5\%].
The GPUs were partly supported by the National IT Industry Promotion Agency (NIPA)'s high-performance computing support program in 2025.

\section{Generative AI Use Disclosure}
Generative AI tools were used only for language polishing and grammar correction.

\bibliographystyle{IEEEtran}
\bibliography{mybib}

@article{Hinton06,
author = {Hinton, Geoffrey E. and Osindero, Simon and Teh, Yee Whye},
journal = {Neural Computation},
pages = {1527--1554},
title = {A Fast Learning Algorithm for Deep Belief Nets},
volume = {18},
year = {2006}
}

@inproceedings{superb,
  title     = {SUPERB: Speech Processing Universal PERformance Benchmark},
  author    = {Yang, Shu-wen and Po-Han Chi and Yung-Sung Chuang and Cheng-I Jeff Lai and Kushal Lakhotia and Yist Y. Lin and Andy T. Liu and Jiatong Shi and Xuankai Chang and Guan-Ting Lin and Tzu-Hsien Huang and Wei-Cheng Tseng and Ko-tik Lee and Da-Rong Liu and Zili Huang and Shuyan Dong and Shang-Wen Li and Shinji Watanabe and Abdelrahman Mohamed and Hung-yi Lee},
  year      = {2021},
  booktitle = {Interspeech 2021},
  pages     = {1194--1198},
  issn      = {2958-1796},
}

@article{transformer,
  title={Attention is all you need},
  author={Vaswani, Ashish and Shazeer, Noam and Parmar, Niki and Uszkoreit, Jakob and Jones, Llion and Gomez, Aidan N and Kaiser, {\L}ukasz and Polosukhin, Illia},
  journal={Advances in neural information processing systems},
  volume={30},
  year={2017}
}

@article{hubert,
  title={Hubert: Self-supervised speech representation learning by masked prediction of hidden units},
  author={Hsu, Wei-Ning and Bolte, Benjamin and Tsai, Yao-Hung Hubert and Lakhotia, Kushal and Salakhutdinov, Ruslan and Mohamed, Abdelrahman},
  journal={IEEE/ACM transactions on audio, speech, and language processing},
  volume={29},
  pages={3451--3460},
  year={2021},
  publisher={IEEE}
}

@article{wavlm,
  title={Wavlm: Large-scale self-supervised pre-training for full stack speech processing},
  author={Chen, Sanyuan and Wang, Chengyi and Chen, Zhengyang and Wu, Yu and Liu, Shujie and Chen, Zhuo and Li, Jinyu and Kanda, Naoyuki and Yoshioka, Takuya and Xiao, Xiong and others},
  journal={IEEE Journal of Selected Topics in Signal Processing},
  volume={16},
  number={6},
  pages={1505--1518},
  year={2022},
  publisher={IEEE}
}

@inproceedings{distilhubert,
  title={Distilhubert: Speech representation learning by layer-wise distillation of hidden-unit bert},
  author={Chang, Heng-Jui and Yang, Shu-wen and Lee, Hung-yi},
  booktitle={ICASSP 2022-2022 IEEE International Conference on Acoustics, Speech and Signal Processing (ICASSP)},
  pages={7087--7091},
  year={2022},
  organization={IEEE}
}

@inproceedings{deepwide,
  title     = {Deep versus Wide: An Analysis of Student Architectures for Task-Agnostic Knowledge Distillation of Self-Supervised Speech Models},
  author    = {Takanori Ashihara and Takafumi Moriya and Kohei Matsuura and Tomohiro Tanaka},
  year      = {2022},
  booktitle = {Interspeech 2022},
  pages     = {411--415},
  issn      = {2958-1796},
}

@inproceedings{fithubert,
  title     = {FitHuBERT: Going Thinner and Deeper for Knowledge Distillation of Speech Self-Supervised Models},
  author    = {Yeonghyeon Lee and Kangwook Jang and Jahyun Goo and Youngmoon Jung and Hoi Rin Kim},
  year      = {2022},
  booktitle = {Interspeech 2022},
  pages     = {3588--3592},
  issn      = {2958-1796},
}

@inproceedings{armhubert,
  title     = {Recycle-and-Distill: Universal Compression Strategy for Transformer-based Speech SSL Models with Attention Map Reusing and Masking Distillation},
  author    = {Kangwook Jang and Sungnyun Kim and Se-Young Yun and Hoirin Kim},
  year      = {2023},
  booktitle = {Interspeech 2023},
  pages     = {316--320},
  doi       = {10.21437/Interspeech.2023-1329},
  issn      = {2958-1796},
}

@inproceedings{
jiang2025tracing,
title={Tracing Representation Progression: Analyzing and Enhancing Layer-Wise Similarity},
author={Jiachen Jiang and Jinxin Zhou and Zhihui Zhu},
booktitle={The Thirteenth International Conference on Learning Representations},
year={2025}
}

@inproceedings{panayotov2015librispeech,
  title={Librispeech: an asr corpus based on public domain audio books},
  author={Panayotov, Vassil and Chen, Guoguo and Povey, Daniel and Khudanpur, Sanjeev},
  booktitle={2015 IEEE international conference on acoustics, speech and signal processing (ICASSP)},
  pages={5206--5210},
  year={2015},
  organization={IEEE}
}

@inproceedings{dphubert,
  title     = {DPHuBERT: Joint Distillation and Pruning of Self-Supervised Speech Models},
  author    = {Yifan Peng and Yui Sudo and Shakeel Muhammad and Shinji Watanabe},
  year      = {2023},
  booktitle = {Interspeech 2023},
  pages     = {62--66},
  doi       = {10.21437/Interspeech.2023-1213},
  issn      = {2958-1796},
}

@inproceedings{jang2024star,
  title={STaR: Distilling Speech Temporal Relation for Lightweight Speech Self-Supervised Learning Models},
  author={Jang, Kangwook and Kim, Sungnyun and Kim, Hoirin},
  booktitle={ICASSP 2024-2024 IEEE International Conference on Acoustics, Speech and Signal Processing (ICASSP)},
  pages={10721--10725},
  year={2024},
  organization={IEEE}
}

@inproceedings{el2025comprehensive,
  title={Comprehensive layer-wise analysis of ssl models for audio deepfake detection},
  author={El Kheir, Yassine and Samih, Younes and Maharjan, Suraj and Polzehl, Tim and M{\"o}ller, Sebastian},
  booktitle={Findings of the Association for Computational Linguistics: NAACL 2025},
  pages={4070--4082},
  year={2025}
}

@inproceedings{gradstack,
  title={Efficient training of language models using few-shot learning},
  author={Reddi, Sashank J and Miryoosefi, Sobhan and Karp, Stefani and Krishnan, Shankar and Kale, Satyen and Kim, Seungyeon and Kumar, Sanjiv},
  booktitle={International Conference on Machine Learning},
  pages={14553--14568},
  year={2023},
  organization={PMLR}
}

@article{midas,
  title={On the inductive bias of stacking towards improving reasoning},
  author={Saunshi, Nikunj and Karp, Stefani and Krishnan, Shankar and Miryoosefi, Sobhan and Jakkam Reddi, Sashank and Kumar, Sanjiv},
  journal={Advances in Neural Information Processing Systems},
  volume={37},
  pages={71437--71464},
  year={2024}
}

@inproceedings{pasad2021layer,
  title={Layer-wise analysis of a self-supervised speech representation model},
  author={Pasad, Ankita and Chou, Ju-Chieh and Livescu, Karen},
  booktitle={2021 IEEE Automatic Speech Recognition and Understanding Workshop (ASRU)},
  pages={914--921},
  year={2021},
  organization={IEEE}
}

@inproceedings{el2024speech,
  title={Speech representation analysis based on inter-and intra-model similarities},
  author={El Kheir, Yassine and Ali, Ahmed and Chowdhury, Shammur Absar},
  booktitle={2024 IEEE International Conference on Acoustics, Speech, and Signal Processing Workshops (ICASSPW)},
  pages={848--852},
  year={2024},
  organization={IEEE}
}

@inproceedings{
nguyen2021do,
title={Do Wide and Deep Networks Learn the Same Things? Uncovering How Neural Network Representations Vary with Width and Depth},
author={Thao Nguyen and Maithra Raghu and Simon Kornblith},
booktitle={International Conference on Learning Representations},
year={2021}
}

@inproceedings{menon2021statistical,
  title={A statistical perspective on distillation},
  author={Menon, Aditya K and Rawat, Ankit Singh and Reddi, Sashank and Kim, Seungyeon and Kumar, Sanjiv},
  booktitle={International Conference on Machine Learning},
  pages={7632--7642},
  year={2021},
  organization={PMLR}
}

@inproceedings{Romero15-iclr,
  author    = {Adriana Romero and
               Nicolas Ballas and
               Samira Ebrahimi Kahou and
               Antoine Chassang and
               Carlo Gatta and
               Yoshua Bengio},
  title     = {FitNets: Hints for Thin Deep Nets},
  booktitle   = {International Conference on Learning Representations},
  year = {2015}
}

@inproceedings{
loshchilov2018decoupled,
title={Decoupled Weight Decay Regularization},
author={Ilya Loshchilov and Frank Hutter},
booktitle={International Conference on Learning Representations},
year={2019}
}

@inproceedings{dicehubert,
  title     = {{DiceHuBERT: Distilling HuBERT with a Self-Supervised Learning Objective}},
  author    = {Hyung-gun Chi and Zakaria Aldeneh and Tatiana Likhomanenko and Oggi Rudovic and Takuya Higuchi and Li-Wei Chen and Shinji Watanabe and Ahmed Hussen Abdelaziz},
  year      = {2025},
  booktitle = {{Interspeech 2025}},
  pages     = {1218--1222},
  doi       = {10.21437/Interspeech.2025-29},
  issn      = {2958-1796},
}

@inproceedings{lin2022smaller,
  title={Is Smaller Always Faster? Tradeoffs in Compressing Self-Supervised Speech Transformers},
  author={Lin, Tzu-Quan and Yang, Tsung-Huan and Chang, Chun-Yao and Chen, Kuang-Ming and Feng, Tzu-hsun and Lee, Hung-yi and Tang, Hao},
  booktitle={2025 IEEE Automatic Speech Recognition and Understanding Workshop (ASRU)},
  year={2025},
  organization={IEEE}
}

\end{document}